\documentclass[%
reprint,
superscriptaddress,
nobibnotes,
 amsmath,amssymb,
 aps,
prb,
]{revtex4-2}

\usepackage{graphicx}
\usepackage{dcolumn}
\usepackage{bm}
\usepackage{hyperref}
\hypersetup{colorlinks=true,
    linkcolor=blue,
    breaklinks=false,
    citecolor=cyan}

\begin{document}

\preprint{APS/123-QED}

\title{Superconductor-Insulator transition in a two-orbital attractive Hubbard model with Hund's exchange}

\author{Laura Torchia}
 \email{ltorchia@sissa.it}
 \affiliation{International School for Advanced Studies (SISSA), via Bonomea 265, 34136 Trieste, Italy}
\author{Massimo Capone}
 \email{capone@sissa.it}
 \affiliation{International School for Advanced Studies (SISSA), via Bonomea 265, 34136 Trieste, Italy}
 \affiliation{Istituto Officina dei Materiali (CNR-IOM), via Bonomea 265, 34136 Trieste, Italy}

\date{\today}

\begin{abstract}
We study a two-orbital attractive Hubbard model with a repulsive Hund's exchange coupling $J$ as an idealized model for a two-band superconductor. This framework is motivated by a system where a large isotropic electron-phonon coupling drives the on-site Hubbard repulsion $U$ negative, while leaving the exchange term unaffected. We focus on the intra-orbital (intra-band) singlet superconducting phase and we solve the model at zero temperature and half-filling using Dynamical Mean-Field Theory.
Already at $J=0$, the two-orbital model features a superconductor-insulator transition as $\vert U\vert$ grows, as opposed to the single-orbital case which remains superconducting for any $U < 0$. We find that a finite $J$ strengthens the effect of the attractive $U$, both in the normal state and, more significantly, in the superconducting state. However, this pushes the system towards an effectively stronger coupling, hence to a faster transition to the insulating state.

Remarkably, the transition from a superconductor to an insulator occurs with a vanishing quasi-particle weight $Z$ and in a scenario that recalls strongly correlated superconductivity close to a Mott transition, even though the present model is dominated by attractive interactions.

\end{abstract}

\maketitle

\section{\label{sec:level1} Introduction}
The theoretical activity on strongly correlated electrons has been long dominated by single-band models where the only local degree of freedom is the electron spin. In the last decade it became clear that these systems should be considered as a specific case in a more general multi-orbital (or multi-component) framework. This change in perspective is motivated by correlated solids with M relevant orbitals and 2M fermionic components per site \cite{Georges_Hund} and system of ultracold atoms with SU(N) symmetry \cite{Gorshkov2010,Pagano2014,Tusi2022}, where $N > 2$ equivalent fermionic components are available for quantum simulation. Here N, which is associated with the nuclear spin, can be as large as 6 for Yb or 10 for Sr.

In the solid-state context, a very popular modeling of the local interactions contains two parameters, the Hubbard interaction $U$, which is the local version of the Coulomb repulsion, and an exchange Hund's coupling $J$, which favours local high-spin configurations. The theoretical activity has so far been largely limited to positive values of $U$ and $J$, which are a direct consequence of the repulsive Coulomb interaction. However, the coupling with phonons can lead to effective negative interactions that can effectively reverse the change of $U$ or $J$ \cite{Capone_Science,Capone2010,Capone2009}.

While the positive-$U$ regime of multiorbital models has been widely studied, much less is known about multi-orbital models with negative charge interactions ($U < 0$) \cite{Koga2015,Zujev_2014,Koga_2011}. 
In this work we move in this direction by considering a minimal model for a two-band superconductor, in which we consider two independent (and, for simplicity, identical) bands coupled by a multi-orbital attraction $-\vert U\vert$ and a positive $J$. 
Our main focus will be to understand how the presence of intra-band coupling influences the superconducting properties of the individual bands. For this reason, we focus  exclusively on intra-orbital (intra-band) pairing, neglecting inter-orbital superconductivity, which is expected to be strongly suppressed in more realistic situations, where significative differences between the two bands disfavour the formation of inter-band pairs. 

Our model can be viewed as a simplified description of a system in which the coupling with a phonon mode inverts the sign of the Hubbard interaction. In this regard, as we argue below, we can connect it to some trends observed in iron-based superconductors (IBS). 

This class of superconductors, second only to the cuprates in terms of ambient-pressure critical temperature, has attracted significant interest since their discovery in 2008 \cite{IBS2008}. Several key features have been identified, including the intrinsic multi-orbital nature of the electronic states\cite{Fernandes2022,Si2023} and the key role of the Hund's exchange coupling \cite{Haule_Kotliar}, which drives an unconventional correlated metal, often referred to as Hund's metal \cite{Haule_Kotliar,deMedici2011,Georges2024}, characterised by orbital-selective correlations \cite{Yu2021,De_Medici_Selective_Mott}. The relationship between the Hund's metal and superconductivity has not been studied with equal depth, although some important results have been reported \cite{Nourafkan2016,Ido2023}. For instance, it has been shown that a Hund's correlated system can coexist with boson-mediated superconductivity \cite{Fanfarillo2020}, and a recent Variational Monte Carlo study found that Hund's coupling and orbital-selective correlations can favour superconductivity  \cite{Vito}. 

On the other hand, although a purely phononic superconducting mechanism has been ruled out early \cite{Lilia}, a significant enhancement of electron-phonon coupling has been reported experimentally for iron-selenium (FeSe) on SrTiO$_3$\cite{Zhang_2017}, where the critical temperature increases from 8K in bulk FeSe to 65K \cite{Zhang_2015}. Theoretical studies have shown a strong pressure dependence of the electron-phonon coupling in FeSe \cite{Mandal2014}. A tentative correlation between superconducting critical temperature and electron-phonon coupling is also observed in heterostructures of iron-arsenic planes and oxides \cite{PhysRevLett.119.107003}.
These observations suggest that the investigation of superconductivity in a multi-orbital model with negative $U$ and positive $J$ is an important piece of information that can help in the search for the mechanisms for superconductivity in IBS and in other Hund's correlated materials.

Another motivation for keeping the model simple is the perspective of reproducing it in quantum simulators based on cold-atom systems. In particular, our model could be realised using a Fano-Feshbach resonance to tune and invert the sign of the interaction in a multi-component system.

The paper is organized as follows. In Sec.\ref{model} we present the model and the method of solution. In Sec.\ref{simple_model} we report our results for a simplified two-orbital model while Sec.\ref{hund_section} is dedicated to the results of our attractive Hubbard model with Hund's coupling. Finally, Sec.\ref{conclusions} contains final conclusions and remarks.

\section{Model and Interaction} 
\label{model}
We consider a two-orbital Hubbard model with Hund's correlations at half-filling. We limit ourselves to the density-density version of the Kanamori Hamiltonian which reads 
 \begin{equation} \label{H}
     \begin{split}
          H =& -\sum_{\langle ij \rangle, m, \sigma} t_m c^{\dag}_{im\sigma} c_{jm\sigma} + U \sum_{i,m} n_{im \uparrow} n_{im\downarrow} \\
          &+  U' \sum_{i,m\neq m'} n_{im\uparrow} n_{im'\downarrow} \\
          &+ (U' -J)\sum_{\substack
          {i,\sigma ,\\m<m'}}n_{im\sigma} n_{im'\sigma}.
     \end{split}
 \end{equation}
where $\langle ij \rangle$ denotes the summations over nearest neighbours, $ c^{(\dagger)}_{im\sigma} $ is the annihilation (creation) operator of an electron on site i, orbital $m=1,2$, with spin $\sigma$, $n_{m\sigma}$ is the number operator, $U$ ($U'$) is the intra-orbital (inter-orbital) local Coulomb interaction, and $J$ is the Hund coupling.  We take for simplicity diagonal hopping $t_{1}=t_{2}=t$. This obviously corresponds to two overlapping Fermi surfaces for the two bands. 

Our target model has $U' = U -2J$ to enforce orbital rotation invariance, and we assume a negative (attractive) $U < 0$ and a positive $J > 0$. If $J=0$ the model becomes a SU(4) Hubbard model.
However in Sec.\ref{simple_model} we will treat $U'$ as an independent negative parameter to describe the evolution from two decoupled single-orbital models to the SU(4) symmetric two-orbital system.

Turning to the microscopic motivation of this model, although $U$ and $J$ are naturally positive if we consider them as arising from the Coulomb interaction, they can turn negative (i.e. attractive) as a result of the electron-phonon interaction. Indeed, if we consider an electron-phonon interaction which couples local phonon modes with fermionic operators $O_i$
\begin{equation}
    H_{\text{el.-ph.}} = g \sum_i O_i ( a^{\dag}_i + a_i ).
\end{equation}
and we integrate out the phonons, we obtain a retarded interaction 
\begin{equation}
    U(\omega)^{\text{ph}} =  - \frac{2g^2\omega_0}{\omega_0^2-\omega^2} \sum_i O_i^2.
    \label{effective}
\end{equation}
Essentially, we have negative (attractive) terms that compete with positive (repulsive) terms originating from purely electronic interactions. In the anti-adiabatic limit of large phonon frequency $\omega_0 \gg t$, we can neglect the frequency $\omega$ in the denominator, and the interaction $U^{\text{ph}}$ in Eq. (\ref{effective}) becomes instantaneous. 
If $O_i = \sum_{m\sigma}  n_{im\sigma} \equiv n_i$, which means that the phonons are coupled with the total local density, the effective interaction is a negative $U$ term which leads to a total effective interaction $U_{\text{eff}} = U - \frac{2g^2}{\omega_0}$. As a result, if the phonon contribution is larger in magnitude, we obtain our model with negative $U$. Analogously, a negative $J$ arises from different phonon modes \cite{Capone2001,Capone_Science,Scazzola2023}, as actually realized in alkali-doped fullerides \cite{Capone2009,Nomura_2015}.

Admittedly, a situation in which the effective interaction becomes negative is not easily realized in strongly correlated systems, where $U$ is by definition sizable. Yet, (i) in Hund's metals the role of $U$ is considered less crucial than that of $J$ and (ii)  retardation effects (neglected in the antiadiabatic limit) can help the electron-phonon driven attraction to prevail over the Coulomb interaction \cite{Sangiovanni2006}.
Thus, our model can be seen as a minimal description of a system where an isotropic electron-phonon coupling prevails over the Hubbard repulsion and leads to phonon-mediated superconductivity while the Hund's term is not affected by phonons.

The main focus of this work is to understand how a multi-orbital superconductor is affected by the Hund's exchange, assuming that the local physics is dominated by an attraction, that we can assume as arising from phonons or from other sources.

As common practice in the field, we discard any possible charge ordering of the system and we study the homogeneous paramagnetic solution, despite the half-filled system on a bipartite lattice is expected to order in the charge channel.
The choice is meant to focus both on the general mechanism for the onset of superconductivity, which is not to be specific to half-filling, and on the intrinsic correlation effects which occur regardless of the nature of the lattice. In the same spirit we do not consider inter-orbital superconductivity, as mentioned above. 

\subsection{Method}
To solve the model and investigate the stability of s-wave superconductivity, we use DMFT \cite{Georges} at zero temperature. DMFT maps the lattice model onto an impurity model, which is solved self-consistently requiring that the impurity Green’s functions coincides with the local component of the lattice one. This method captures dynamical correlations and treats interactions nonperturbatively, providing a more quantitative description of superconductivity than static BCS mean-field theory. It has been successfully applied to study the Mott transition, strongly correlated metals, and superconducting and superfluid states in correlated fermionic systems.
The impurity model is solved at T=0 using Lanczos/Arnoldi exact diagonalization (ED) \cite{Caffarel1994,Capone2007,AMARICCI2022108261}. We employ an artificial inverse temperature $\beta=400$ to define the Matsubara frequency grid. We also test our results for different values of $\beta$ and confirm their independence on the grid.

In our case, the Hamiltonian (\ref{H}) is mapped onto a an Anderson model with attractive coupling where the bath also contains superconducting parameters. We considered $N_b=4$ bath sites, each with 2 orbitals, and we have verified that the dependence of our results on the finite bath size is negligible. The effect of the discretized bath is essentially only visible in real-frequency spectra, as we discuss in the following.
We consider the model on a Bethe lattice with a semicircular density of states  $D(\epsilon)=\frac{2}{\pi D^2}\sqrt{D^2-\epsilon^2}$, whose half width $D$ is used as the energy unit. This is a very popular choice in DMFT as it represents an infinite-coordination lattice, where DMFT becomes exact, yet it has a finite bandwidth, which makes a closer connection with finite-dimensional lattices.
We consider the system at global half-filling $\sum_{m\sigma} n_{m\sigma}=2$, where $n_{m\sigma} = \langle c^{\dagger}_{im\sigma}c_{im\sigma}\rangle$ on any lattice site i (given that our solutions are translationally invariant).

Superconductivity is conveniently treated in the Nambu-Gor'kov formalism where the Green's function is
a 2$\times$2 matrix. For singlet pairing it can be written as 
\begin{equation}
    \begin{split}
        \hat{G}(\boldsymbol k, \tau)&\equiv \langle{T \Psi_{\boldsymbol k}(\tau)\Psi^\dagger_{\boldsymbol k}(0) \rangle}\\
        &=\begin{pmatrix}
            G(\boldsymbol k, \tau)& \mathcal{F}(\boldsymbol k, \tau)\\
            \mathcal{F}(\boldsymbol k, \tau)^* & -G(\boldsymbol -k, -\tau)
        \end{pmatrix}.
    \end{split}
\end{equation}
Here, $G$ and $F$ represent the normal and anomalous components of the lattice Green's function, respectively.
The same matrix structure holds for the self energy, which in DMFT is momentum independent, with a off-diagonal anomalous component $S$ and a normal diagonal one $\Sigma$:
\begin{equation} \label{Smatrix}
    \begin{split}
        \hat{\Sigma}(i\omega_n)& =\begin{pmatrix}
        \Sigma(i\omega_n)& S( i\omega_n)\\
        S(i\omega_n) &  -\Sigma( -i\omega_n)
    \end{pmatrix}
    \end{split}
\end{equation}
where $\Sigma(-i\omega_n)=\Sigma^*(i\omega_n)$.
For our analysis, it is useful to write out the Green’s function components explicitly
\begin{equation}
    \begin{split}
        & G(\boldsymbol{k},i\omega_n)=\frac{i\omega_n -\epsilon_{\boldsymbol{k}}-\Sigma(i\omega_n)}{|i\omega_n-\Sigma(i\omega_n)-\epsilon_{\boldsymbol{k}}|^2+|S|^2}\\
        & \mathcal{F}(\boldsymbol{k},i\omega_n)=\frac{-S(i\omega_n)}{|i\omega_n-\Sigma(i\omega_n)-\epsilon_{\boldsymbol{k}}|^2+|S|^2}
    \end{split}
\end{equation}

We note that in our two-orbital system, all quantities have an additional orbital structure. However, in the present calculations orbital degeneracy is conserved which implies orbital-independent Green's functions and self-energies (the latter up to phase factor).\\
We characterize completely the superconducting phase in terms of the superconducting order parameter (pairing amplitude) $\phi$
\begin{equation}
    \phi \equiv \phi_m =  \langle{d_{im\uparrow}d_{im\downarrow}\rangle} = \lim_{\tau\rightarrow 0_{+}}\mathcal{F}_m(\tau),
\end{equation}
the superfluid stiffness $D_s$, that measures the coherence of the superconducting state and its rigidity to fluctuations of the phase of the order parameter
\begin{equation}
    D_s= \frac{4e^2\pi}{\hbar}k_BT\sum_{i\omega_n}\int d\epsilon D(\epsilon)V(\epsilon)|\mathcal{F}(\boldsymbol{k},i\omega_n)|^2,
\end{equation}
and the quasi-particle weight $Z$, that gives us an indication of the single-particle coherence of the system
\begin{equation}
    Z=\biggl(1-\frac {\partial\mathfrak{Im}\Sigma (\boldsymbol k_F,i\omega_n)}{\partial i\omega_n}\bigg\rvert_0\biggr)^{-1}.
\end{equation}
This observable is commonly used to track the transition to a correlation-driven insulating state, where $Z \rightarrow 0$ due to a divergence of $\mathfrak{Im}\,\Sigma(\boldsymbol{k}_F, i\omega_n)$. However, as expected, we observed a light dependence on $\beta$ near the transition. Therefore, we also identify the transition point by observing discontinuities in correlation functions, accompanied by a divergence in the self-energy.

\section{From one to two orbitals}
\label{simple_model}
Before turning to the full model Eq. (\ref{H}), we solve a simplified model that highlights a crucial difference between single- and two-orbital attractive models and the role of inter-orbital interactions:
\begin{equation} \label{Hsimple}
     \begin{split}
          H =& -t \;\sum_{\langle ij \rangle, m, \sigma} \;c^{\dag}_{im\sigma} c_{jm\sigma} - \vert U\vert\sum_{i,m} n_{im \uparrow} n_{im\downarrow} \\
          &-\vert U'\vert \sum_{i,m\neq m'} n_{im\uparrow} n_{im'\downarrow} \\
          &-\vert U'\vert \sum_{\substack
          {i,\sigma ,\\m<m'}}n_{im\sigma} n_{im'\sigma}.
     \end{split}
 \end{equation}

where we neglect the $J$ term and we treat the interorbital interaction as an independent attractive coupling (i.e. we do not assume $U' = U -2J$). 

In this way we can connect the limit of two decoupled single-orbital attractive Hubbard models ($U'=0$) with the fully SU(4) model for $U' = U$. 

We recall the well-established physics of the single-band attractive Hubbard model, where increasing the interaction magnitude $|U|$ drives a smooth crossover from a BCS-like Cooper pair regime to a BEC-like superconducting state formed by composite bosons \cite{Toschi_BCSBEC,Kyung2006}. As $\vert U\vert$ increases the modulus of the order parameter $\phi$ increases, even if the critical temperature (not computed here) decreases as a result of the reduced coherence of the tightly bound pairs \cite{Toschi_2005,Toschi_BCSBEC,Kyung2006}. The quasi-particle weight is instead finite and close to 1 for every value of $U$ with a shallow minimum for intermediate interactions, where the crossover takes place.
\begin{figure}[t!]
    \centering
    \includegraphics[width=1.0\linewidth]{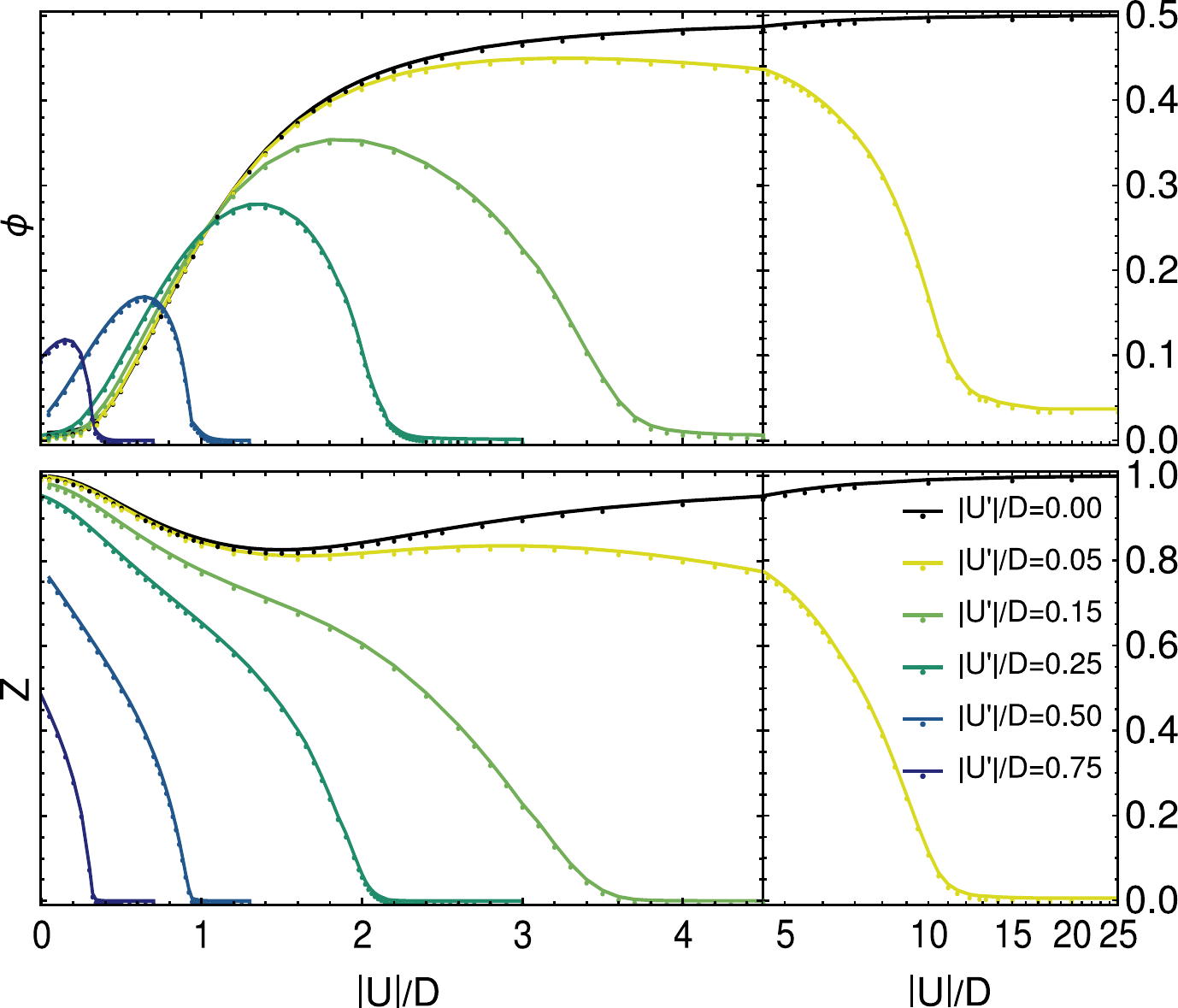}
    \caption{Superconducting order parameter $\phi$ (top) and quasi-particle weight $Z$ (bottom) as functions of the effective local interaction $|U|/D$ for different values of the inter-orbital coupling $|U'|/D$. Both $U/D$ and $U'/D$ are considered attractive. The black lines represent the case of two decoupled orbitals, while increasing color gradient corresponds to increasing $U'/D$: from weak coupling (yellow) to stronger inter-orbital coupling (dark blue). }
    \label{fig:phi_int}
\end{figure}

In Fig. \ref{fig:phi_int} we report the evolution of $\phi$ and $Z$ as a function of $U$ for different values of $U'$. The most spectacular change emerges on the strong-coupling side, where, as soon as $U'$ is switched on, a rapid drop of $\phi$ takes place for large $U$, accompanied by a vanishing of $Z$. For the smallest value we considered, $U' = 0.05U$, we find a finite (yet very small) plateau for $\phi$ in strong coupling, while the minimum of $Z$, roughly associated with the BCS-BEC crossover, is still present. Increasing $|U'|$ we find that $\phi$ goes to zero at large $U$ and $Z$ drops to zero without any intermediate-coupling anomaly.

The vanishing of $Z$ implies an insulating solution, hence the BCS-BEC crossover is replaced by a transition between a superconductor and a pairing insulator, a collection of localized pairs without coherence. In the single-band model this phase can only be stablized at finite temperature or inhibiting artificially the pairing, while here it prevails over the BEC superconductor.  We thus find that the inclusion of an attractive inter-orbital interaction proves detrimental for superconductivity in the strong coupling regime. On the other hand, on the weak coupling site, switching on $U'$ induces a fine order parameter already for $U=0$.  

We note that similar results have been obtained in Ref.\cite{Koga2015} for a model with negative $U$ and positive $U'$, which is indeed mapped onto the present model via a particle-hole transformation on one orbital.

We can indeed gain understand the destruction of the superconducting state induced by $U'$ through a simple argument. Starting from two attractive single-band models, a finite negative $U'$ introduces correlations between the two orbitals, favoring configurations where the intra-orbital pairs are on the same site. 
This means that the pairs in the two orbitals are bound and they move together. In strong coupling, this makes an important difference: while the pairs of a single-band model move via second-order processes, with an effective hopping $\propto t^2/U$, in order to move two bound pairs, one has to move four electrons, leading to a much smaller $t^4/U^3$ effective hopping. This reduces substantially the coherence in the strong-coupling limit, leading to the possibility that the insulating pairing state prevails. 

This argument is expected to be quite general. As we will see, it essentially holds also for our full model including the Hund's coupling.

\section{Effect of Hund's coupling}
\label{hund_section}
\subsection{Normal phase}

\begin{figure}[t!]
    \centering
        \includegraphics[width=\linewidth]{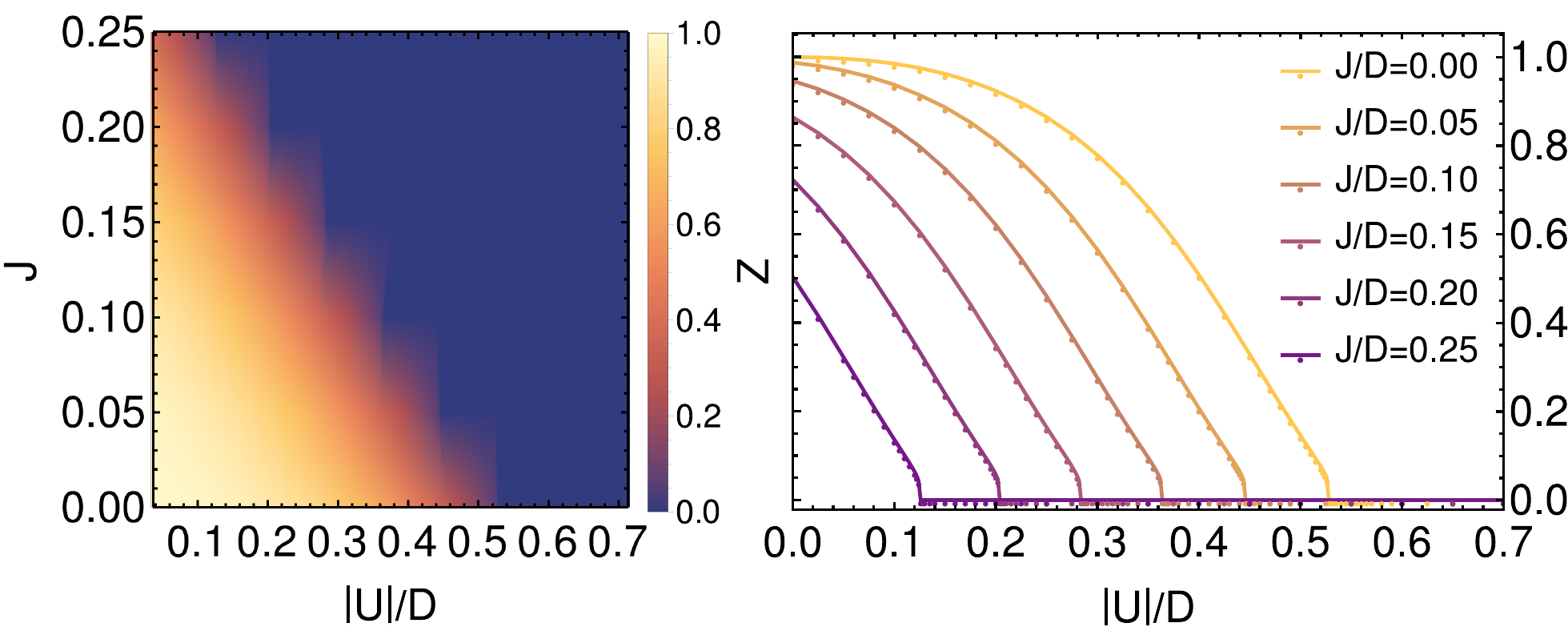}
    \caption{
     (left) Color plot of the quasi-particle weight as function of the Hund's coupling $J/D$ and the effective local interaction $U/D$. Lighter areas indicates higher degree of metallicity, i.e. $Z$ closer to 1, while in darker areas the system is approaching the insulating state, i.e. $Z$ goes to 0.
     (right) Cuts at fixed $J$ as a function of $U/D$. The color gradient scale corresponds to increasing values of the Hund's coupling $J$: from $J=0$ (yellow) to $J=0.25$ (dark purple).     
    }
    \label{fig:zplotN}
\end{figure}
We begin our study of the model Eq.(\ref{H}) by examining the normal phase solution in which we inhibit any type of broken symmetry, including superconductivity. In the single-band attractive model, the normal state features a transition from a metal to a pairing insulator at intermediate $U$, which is the attractive counterpart of the Mott transition \cite{Keller2001,Capone2002,Toschi_2005}.

In Fig.(\ref{fig:zplotN}) we show the quasi-particle weight $Z$ as a function of $U$ for different values of $J$. The monotonic decrease of $Z$ signals the progressive destruction of the metallic phase up to a critical $U$ where $Z$ vanishes leading to an insulating phase. The critical value for the Mott transition $\vert U_c \vert$ is already relatively small for $J =0$ and it is clear that increasing $J$ reduces $\vert U_c \vert$.

In order to characterize the evolution towards the metal-insulator transition, we compute density-density correlations $C_{mm'} =\langle{n_{im} n_{im'}\rangle}-\langle{n_{im}}\rangle\langle{n_{im'}}\rangle $ on any lattice site i. The intra-orbital term is proportional to the double occupancy of the orbital $d_m$, namely $C_{mm}=2\cdot d_m$. Due to orbital symmetry, we have that $C_{12}=C_{21}$ and $d_1=d_2$.
The curves show a monotonic behavior, with a clear jump that signals the first order character of the  transition. The limiting results of the two observables clearly confirm the nature of the insulating solution. On one hand, the double occupancy of each orbital reaches the maximum value 0.5, confirming the presence of pairs on both orbitals. $C_{12}$ asymptotically reaches 1, signaling that the two orbitals are completely correlated, namely two pairs are present on the same site in the two orbitals, as we discussed in Sec.\ref{simple_model}. This agrees with the atomic limit $t = 0$ in which we find a solution where half of the sites are filled and each orbital is doubly occupied. 

These calculations confirm therefore that also our complete model at strong coupling reaches the same pairing insulator we obtained for Eq.(\ref{Hsimple}). In the next section we discuss the effect of superconductivity. 

\begin{figure}[t]
    \centering
    \includegraphics[width=\linewidth]{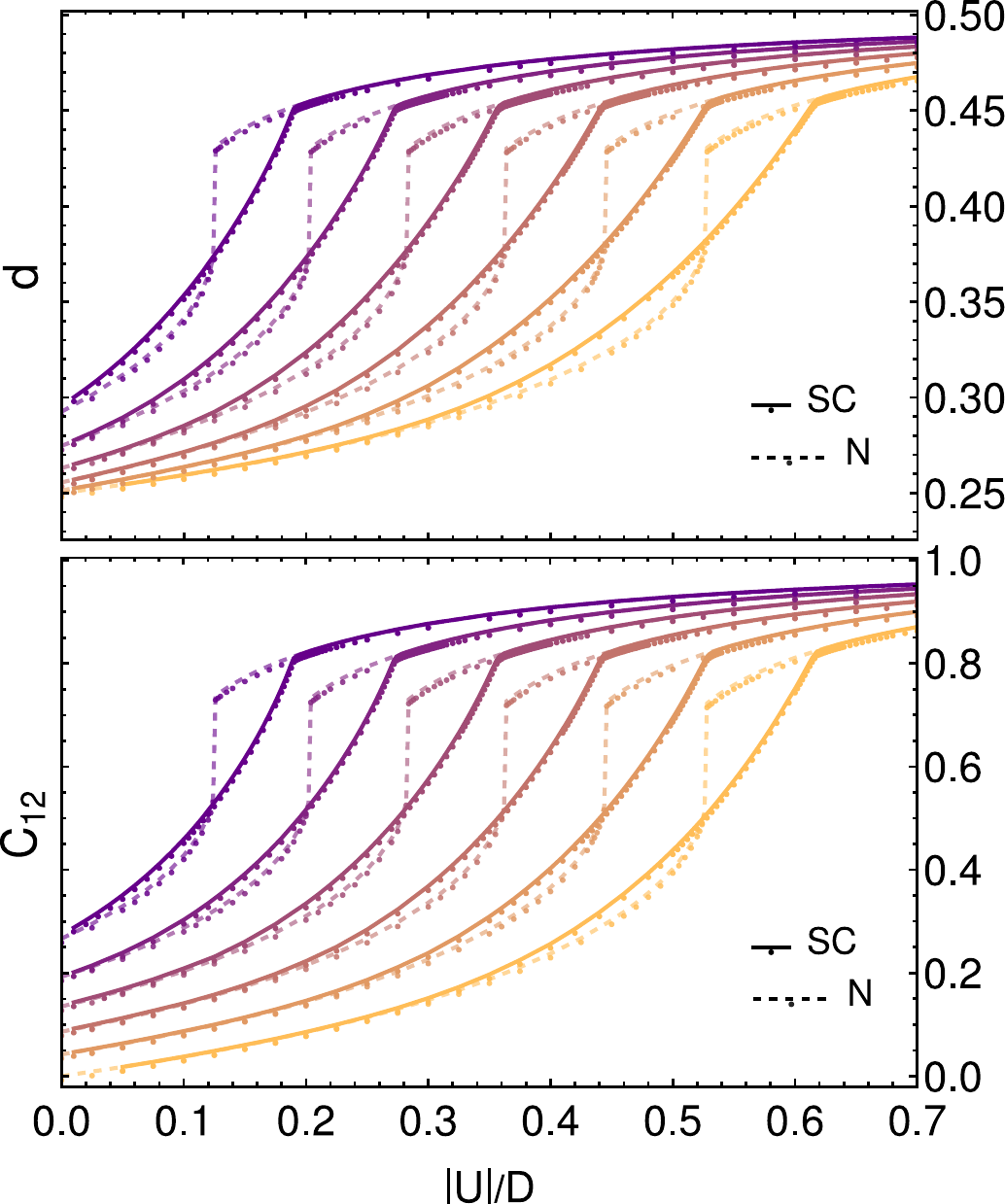}
    \caption{Double occupations (top) and inter-orbital correlations (bottom) as function of the effective local interaction $|U|/D$ for different values of the Hund's coupling, legend in Fig.(\ref{fig:zplotN}). Dashed lines are data in the normal phase, solid lines are the same quantities in the superconducting phase.}
    \label{fig:corr}
\end{figure}

\subsection{Superconducting state}
We now allow for superconductivity, and we compare the correlation functions in the superconducting phase (solid lines) with the normal phase (dotted line) in Fig.(\ref{fig:corr}). While the general trend is similar, we notice that both quantities are slightly larger in the superconducting state, and that, since the superconducting solution is more stable, the transition to the pairing insulator takes place  at higher values of $|U|$. Interestingly, the transition appears continuous, as opposed to the normal-state discontinuous evolution. As a matter of fact the superconducting solution replaces the normal state and smoothly connects with the pairing insulator at $U_c$.

\begin{figure}[t!]
    \centering
    \includegraphics[width=0.95\linewidth]{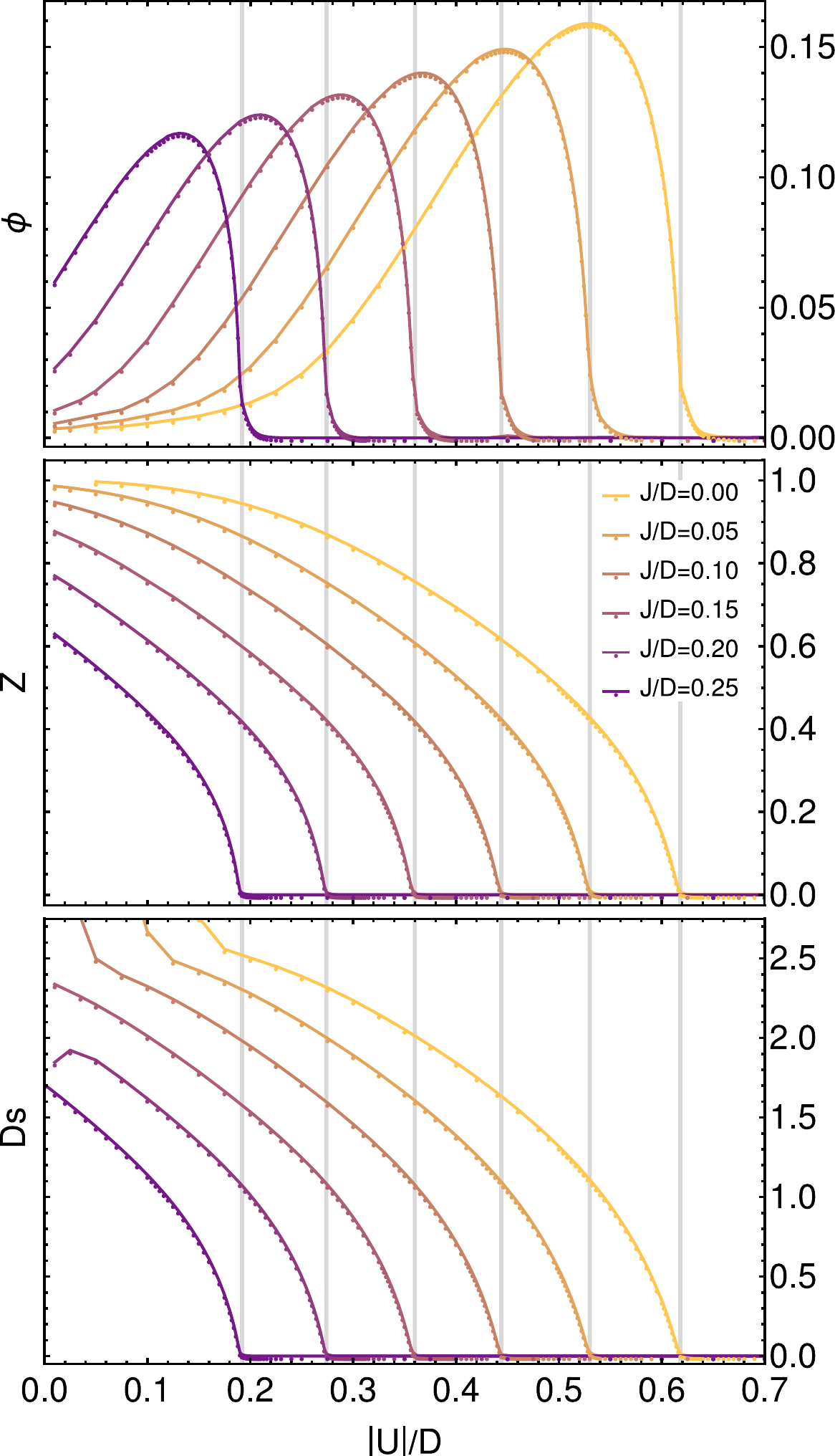
}
    \caption{Superconducting order parameter (top), quasi-particle weight (middle), superfluid stiffness (bottom), as function of the effective local interaction $|U|/D$ for different values of the Hund's coupling. The grey vertical lines signal the transition to the pairing insulting phase.}
    \label{fig:phi_z_Ds}
\end{figure}

We now turn to the characterization of the superconducting state in terms of the order parameter $\phi$, the quasi-particle weight $Z$ and the superfluid stiffness $D_S$ (Fig.(\ref{fig:phi_z_Ds})).

For $J=0$ we recover results for $\phi$ and $Z$ similar to those of Fig.(\ref{fig:phi_int}) for the largest values of $|U'|$.  When $J$ increases we clearly see a shift of the profiles towards lower values of $|U|$ and a decrease of the maximum value of the order parameter. Similar trends are seen in $Z$ and $D_s$
We notice a finite $\phi$ present in the region of low $|U|$. For $J=0$ we ascribe this value to a numerical error associated with the difficulty to resolve the exponentially small BCS-like order parameter. On the other hand, as $J$ increases, it can actually induce a finite intra-orbital order parameter.

\begin{figure}[b]
    \centering
    \includegraphics[width=\linewidth]{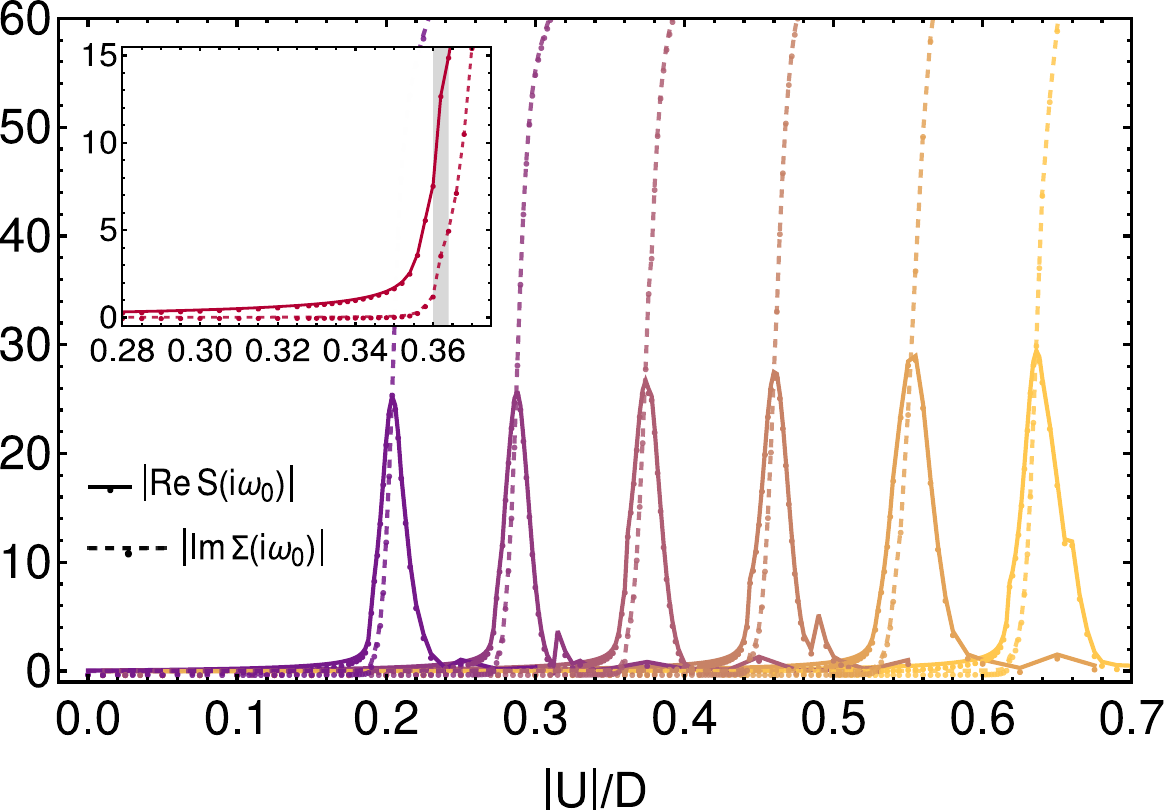}
    \caption{Self-energies (Imaginary part of the normal component and real part of the anomalous component) 
    computed at the first Matsubara frequency $\omega_0$, where $\beta = 400$ in order to highlight the singular behavior at the pairing transition. The functions are shown as function of the effective local interaction $|U|/D$ for different values of the Hund's coupling, legend in Fig.(\ref{fig:phi_z_Ds}).Inset: behaviour close to the transition for $J=0.15$.}
    \label{fig:Sigma_div}
\end{figure}

All the curves of $\phi$ for different values of $J$ exhibit similar behavior: starting from a low value of $\phi$ at $|U|=0$, increasing the coupling raises the order parameter, which reaches its peak just before the superconductor-pairing insulator transition, signaled by the grey bars. Despite being very similar, the different curves can not be superimposed via a shift of $U$ and a rescaling of the peak height. 
At the transition point, both the quasi-particle weight and the superfluid stiffness vanish, leading to an abrupt drop of the order parameter to zero, and the consequent formation of the pairing insulating state. 
We notice that, in terms of the order parameter $\phi$ our results show that for small $\vert U\vert$ the effect of $J$ is to increase the order parameter, while in strong coupling it leads to a reduction of the critical $U$, hence it opposes to superconductivity in favor of the pairing insulator. So, while $J$ favours in general the pairing between the electrons, this turns into an actual increase of the order parameter only at weak $\vert U\vert$, while for large $\vert U\vert$ it favors the localization of the pairs.
This is a totally new scenario which can not be realized in single-band systems, but it requires the presence of different interactions, and in particular of the Hund's coupling.\\
We specify that at weak coupling, $U \rightarrow 0$, and zero or small $J$, the order parameter is proportional to the critical temperature and therefore exponentially small. To achieve better resolution in this regime, we increased the density of the Matsubara frequency grid by increasing $\beta$. However, as we approach the real axis, we also get closer to singularities, which can lead to numerical instabilities. As a result, the stiffness in this region exhibits deviations from the expected behavior.

Through DMFT we have also access to dynamical (frequency dependent) quantities such as self-energies and the spectral functions. We start from a comparison between the normal component of the self-energy, $\Sigma(i\omega_n)$, whose divergence (of the imaginary part) at $i\omega_n \rightarrow 0$ marks the transition to the insulating phase, and the anomalous self-energy $S(i\omega_n)$, which gives us information on the superconducting pairing amplitude and gap.

In Fig. \ref{fig:Sigma_div} we plot the two self-energies at the smallest fermionic Matsubara frequency $\omega_0 = \pi/\beta$. In a metallic state $\mathfrak{Im}\Sigma(i\omega_n)$ vanishes in the limit of zero frequency, and it is very small at $\omega_0$. When we approach an insulating solution $\mathfrak{Im}\Sigma(i\omega_0)$ grows and it can become very large if the coherence scale of the metal becomes smaller than $\omega_0$. When we enter the insulating region $\mathfrak{Im}\Sigma(i\omega_0)$ diverges in the zero-frequency limit, leading to a very large value at $\omega_0$. The behavior of $S(i\omega_0)$ is not expected to display singularities.

The results have a consistent behavior for different values of $J$. In particular, at low couplings far from the transition, where the system preserves its metallicity,   $\mathfrak{Im}\Sigma(i\omega_0)$ is essentially zero (and it would reach zero smoothly in the limit of zero frequency) while the presence of a superconducting state leads to a non-vanishing anomalous self energy. However, approaching the transition, we observe first a very fast growth of $\mathfrak{Re}S(i\omega_0)$, followed by one in $\mathfrak{Im}\Sigma(i\omega_0)$, signaling the approach and the onset of an insulating state.
The inset in Fig.(\ref{fig:Sigma_div}) clearly shows that the increase of the anomalous self-energy is separated from the growth of the normal component. 
At higher interactions, the insulating state prevails, and the normal component crosses the anomalous one, which eventually goes to zero. 
A similar behavior reflects the fact that, close to the transition, a very large bare amplitude is necessary to obtain a finite order parameter in the presence of a large quasi-particle renormalization, since the superconducting component of the spectral gap is given by $ZS(i\omega_n \to 0)$.

The fast growth of the anomalous self-energy explains the increase of the order parameter before the transition that we have discussed. As a matter of fact, the superconducting phase close to the transition to a pairing insulator is characterized by a strong boost due to strong correlations which is completely absent in the single-band model, where insulating phase is not stable at $T=0$ and a smooth maximum of the order parameter takes place at intermediate interactions.

\begin{figure}[t!]
     \centering
     \includegraphics[width=\linewidth]{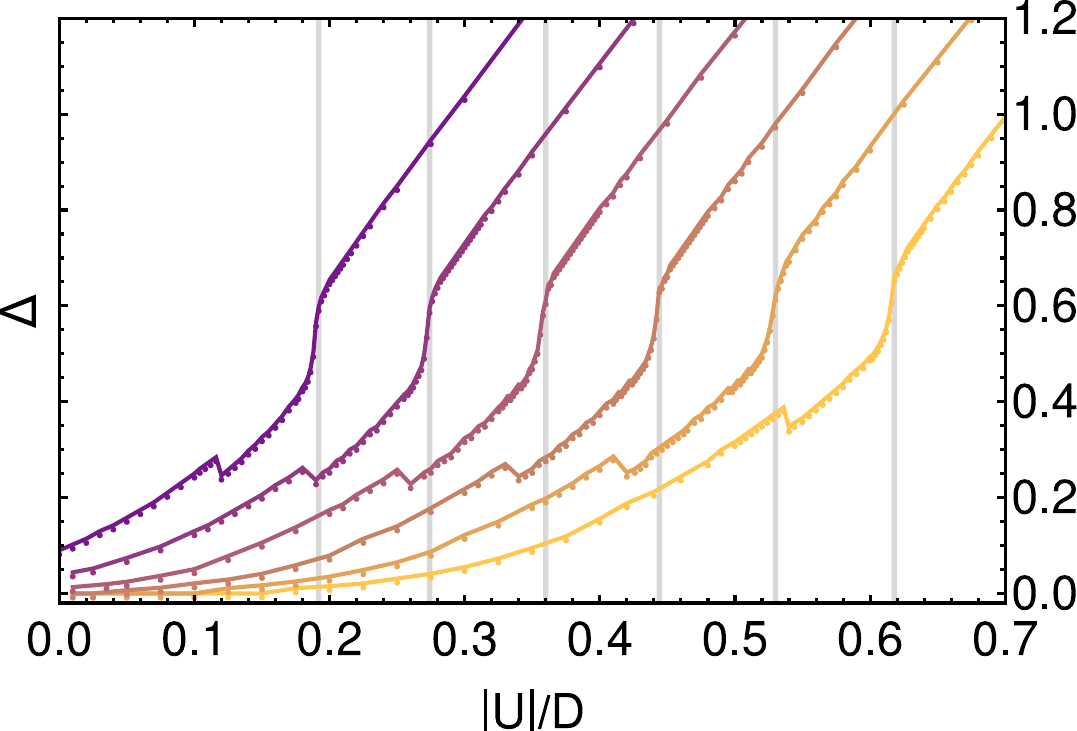}
     \caption{Gap of the impurity spectral function as function of the effective local interaction $|U|/D$ for different values of the Hund's coupling, legend in Fig.(\ref{fig:phi_z_Ds}).}
     \label{fig:gap}
 \end{figure}

The fingerprint of a strongly correlated superconductivity is further confirmed by extracting the gap from the spectral function $A(\omega)=-\frac{1}{\pi}\mathfrak{Im}G_{imp}(\omega)$, displayed in Fig.(\ref{fig:gap}). Indeed, starting from low $|U|$, the spectral gap arises mainly from the superconducting ordering and increases gradually with the module of interaction. Near the superconductor-insulator transition, the gap evolution changes, and the spectral gap is sharply enhanced. Eventually, the preformed boson-like pairs fully localize and the superconducting state loses phase coherence. As $Z$ and $D_S$ approach zero, the system smoothly transitions to a pairing insulating state and the order parameter vanishes. The system enters the insulating phase, and the superconducting gap is replaced by the insulating gap. We clarify that the kink in the gap functions at low coupling is essentially a numerical effect due to the discretization of the bath, which is naturally more visible in real-frequency observables. In the specific case, the small jump follows from a slightly different convergence coming from the superconducting and insulating sides of the transition. However, this is only seen in the spectral function and it does not lead to any discontinuity in any other observable. 

\section{Conclusions}
\label{conclusions}
We have studied using DMFT the intra-orbital superconducting state of a two-orbital Hubbard model described by a density-density Kanamori Hamiltonian with negative $U$ and positive $J$.

We find that, under rather general conditions, the two-orbital model has a profoundly different superconducting phase diagram with respect to the single-orbital attractive Hubbard model.  While the latter model has a superconducting ground state for every value of $U$ and displays a BCS-BEC crossover as $|U|$ is increased, the two-orbital model displays a transition from the superconducting state to a pairing insulator, a collection of localized pairs,  at relatively small values of $|U|$. 

The destruction of the superconducting state leading to the pairing insulator has been found also in a simpler model with only intra- and inter-orbital attractions $U$ and $U'$, where we built a simple physical picture based on the fact that pairs on the two orbitals tend to attract each other on the same site gaining and energy $-|U'|$, leading to four-electron entities, which are much less mobile than standard pairs when $|U|$ is large. 

This leads to a destruction of the BEC regime and to a dome-shape behavior of the superconducting order parameter which goes to zero continuously at the superconductor-insulator transition. As $|U'|$ grows the critical $|U|$ is reduced.

In the Kanamori model, the Hund's exchange coupling $J$  plays a similar role to $U'$: it shifts the superconducting domes to lower values of the effective $|U|$, favoring the superconductor-insulator transition. 
Overall, the effect of Hund's coupling $J$ is to enhance the effects of the attractive $U$, not only in normal state but especially in the superconducting state.
Although we have not included the exchange and pair-hopping terms in the Hamiltonian (\ref{H}), we note that these terms would favor fully rotationally invariant high-spin states, while our model selects only high-spin $S_z$ states. Therefore, we believe they have little impact on the stability of superconductivity.

The analysis of the normal and superconducting self-energies shows that the approach to the pairing transition is very similar to a Mott transition, with a divergent self-energy. The anomalous (superconducting) self-energy is small in the weak-coupling region far from the transition, but it grows very rapidly before the transition, leading to a finite superconducting order parameter. These are characteristic signatures of a strongly correlated superconducting state, that is stabilized close to a metal-insulator transition, as it happens in the fullerides \cite{Capone_Science,Nomura_2015}.

While our model is not intended to describe a specific material, it can be seen as an extreme case of the physics observed in iron-based superconductors when the electron-phonon coupling is boosted, as it happens in mono-layer materials. Our calculations suggest, that in multi-orbital systems of this kind, the conventional BCS-BEC crossover is hardly realized\cite{Witt2024}, even if a dome-shaped behavior of the superconducting parameter is still observed, and that results based on single-orbital models are hardly relevant for multi-band systems with finite inter-orbital interactions. Future work is needed to test if this scenario can actually be realized for realistic values of parameters, where the electron-phonon coupling is smaller than the bare $U$, including retardation effects, which help to overcome the effect of the repulsion.
\\

\begin{acknowledgments}
The authors are thankful to M. Ferraretto and S. Giuli for the useful discussions. We acknowledge financial support of MUR via PRIN 2020 (Prot. 2020JLZ52N 002) programs, and by the European Union - NextGenerationEU through PRIN 2022 (Prot. 20228YCYY7), MUR PNRR Project No. PE0000023-NQSTI, and No. CN00000013-ICSC.
\end{acknowledgments}

\bibliographystyle{apsrev4-2}
\bibliography{biblio,References}

\begin{thebibliography}{41}%
\makeatletter
\providecommand \@ifxundefined [1]{%
 \@ifx{#1\undefined}
}%
\providecommand \@ifnum [1]{%
 \ifnum #1\expandafter \@firstoftwo
 \else \expandafter \@secondoftwo
 \fi
}%
\providecommand \@ifx [1]{%
 \ifx #1\expandafter \@firstoftwo
 \else \expandafter \@secondoftwo
 \fi
}%
\providecommand \natexlab [1]{#1}%
\providecommand \enquote  [1]{``#1''}%
\providecommand \bibnamefont  [1]{#1}%
\providecommand \bibfnamefont [1]{#1}%
\providecommand \citenamefont [1]{#1}%
\providecommand \href@noop [0]{\@secondoftwo}%
\providecommand \href [0]{\begingroup \@sanitize@url \@href}%
\providecommand \@href[1]{\@@startlink{#1}\@@href}%
\providecommand \@@href[1]{\endgroup#1\@@endlink}%
\providecommand \@sanitize@url [0]{\catcode `\\12\catcode `\$12\catcode `\&12\catcode `\#12\catcode `\^12\catcode `\_12\catcode `\%12\relax}%
\providecommand \@@startlink[1]{}%
\providecommand \@@endlink[0]{}%
\providecommand \url  [0]{\begingroup\@sanitize@url \@url }%
\providecommand \@url [1]{\endgroup\@href {#1}{\urlprefix }}%
\providecommand \urlprefix  [0]{URL }%
\providecommand \Eprint [0]{\href }%
\providecommand \doibase [0]{https://doi.org/}%
\providecommand \selectlanguage [0]{\@gobble}%
\providecommand \bibinfo  [0]{\@secondoftwo}%
\providecommand \bibfield  [0]{\@secondoftwo}%
\providecommand \translation [1]{[#1]}%
\providecommand \BibitemOpen [0]{}%
\providecommand \bibitemStop [0]{}%
\providecommand \bibitemNoStop [0]{.\EOS\space}%
\providecommand \EOS [0]{\spacefactor3000\relax}%
\providecommand \BibitemShut  [1]{\csname bibitem#1\endcsname}%
\let\auto@bib@innerbib\@empty
\bibitem [{\citenamefont {Georges}\ \emph {et~al.}(2013)\citenamefont {Georges}, \citenamefont {Medici},\ and\ \citenamefont {Mravlje}}]{Georges_Hund}%
  \BibitemOpen
  \bibfield  {author} {\bibinfo {author} {\bibfnamefont {A.}~\bibnamefont {Georges}}, \bibinfo {author} {\bibfnamefont {L.~d.}\ \bibnamefont {Medici}},\ and\ \bibinfo {author} {\bibfnamefont {J.}~\bibnamefont {Mravlje}},\ }\href {https://doi.org/10.1146/annurev-conmatphys-020911-125045} {\bibfield  {journal} {\bibinfo  {journal} {Annual Review of Condensed Matter Physics}\ }\textbf {\bibinfo {volume} {4}},\ \bibinfo {pages} {137} (\bibinfo {year} {2013})},\ \Eprint {https://arxiv.org/abs/https://doi.org/10.1146/annurev-conmatphys-020911-125045} {https://doi.org/10.1146/annurev-conmatphys-020911-125045} \BibitemShut {NoStop}%
\bibitem [{\citenamefont {Gorshkov}\ \emph {et~al.}(2010)\citenamefont {Gorshkov}, \citenamefont {Hermele}, \citenamefont {Gurarie}, \citenamefont {Xu}, \citenamefont {Julienne}, \citenamefont {Ye}, \citenamefont {Zoller}, \citenamefont {Demler}, \citenamefont {Lukin},\ and\ \citenamefont {Rey}}]{Gorshkov2010}%
  \BibitemOpen
  \bibfield  {author} {\bibinfo {author} {\bibfnamefont {A.~V.}\ \bibnamefont {Gorshkov}}, \bibinfo {author} {\bibfnamefont {M.}~\bibnamefont {Hermele}}, \bibinfo {author} {\bibfnamefont {V.}~\bibnamefont {Gurarie}}, \bibinfo {author} {\bibfnamefont {C.}~\bibnamefont {Xu}}, \bibinfo {author} {\bibfnamefont {P.~S.}\ \bibnamefont {Julienne}}, \bibinfo {author} {\bibfnamefont {J.}~\bibnamefont {Ye}}, \bibinfo {author} {\bibfnamefont {P.}~\bibnamefont {Zoller}}, \bibinfo {author} {\bibfnamefont {E.}~\bibnamefont {Demler}}, \bibinfo {author} {\bibfnamefont {M.~D.}\ \bibnamefont {Lukin}},\ and\ \bibinfo {author} {\bibfnamefont {A.~M.}\ \bibnamefont {Rey}},\ }\href {https://doi.org/10.1038/nphys1535} {\bibfield  {journal} {\bibinfo  {journal} {Nature Physics}\ }\textbf {\bibinfo {volume} {6}},\ \bibinfo {pages} {289} (\bibinfo {year} {2010})}\BibitemShut {NoStop}%
\bibitem [{\citenamefont {Pagano}\ \emph {et~al.}(2014)\citenamefont {Pagano}, \citenamefont {Mancini}, \citenamefont {Cappellini}, \citenamefont {Lombardi}, \citenamefont {Sch{\"a}fer}, \citenamefont {Hu}, \citenamefont {Liu}, \citenamefont {Catani}, \citenamefont {Sias}, \citenamefont {Inguscio},\ and\ \citenamefont {Fallani}}]{Pagano2014}%
  \BibitemOpen
  \bibfield  {author} {\bibinfo {author} {\bibfnamefont {G.}~\bibnamefont {Pagano}}, \bibinfo {author} {\bibfnamefont {M.}~\bibnamefont {Mancini}}, \bibinfo {author} {\bibfnamefont {G.}~\bibnamefont {Cappellini}}, \bibinfo {author} {\bibfnamefont {P.}~\bibnamefont {Lombardi}}, \bibinfo {author} {\bibfnamefont {F.}~\bibnamefont {Sch{\"a}fer}}, \bibinfo {author} {\bibfnamefont {H.}~\bibnamefont {Hu}}, \bibinfo {author} {\bibfnamefont {X.-J.}\ \bibnamefont {Liu}}, \bibinfo {author} {\bibfnamefont {J.}~\bibnamefont {Catani}}, \bibinfo {author} {\bibfnamefont {C.}~\bibnamefont {Sias}}, \bibinfo {author} {\bibfnamefont {M.}~\bibnamefont {Inguscio}},\ and\ \bibinfo {author} {\bibfnamefont {L.}~\bibnamefont {Fallani}},\ }\href {https://doi.org/10.1038/nphys2878} {\bibfield  {journal} {\bibinfo  {journal} {Nature Physics}\ }\textbf {\bibinfo {volume} {10}},\ \bibinfo {pages} {198} (\bibinfo {year} {2014})}\BibitemShut {NoStop}%
\bibitem [{\citenamefont {Tusi}\ \emph {et~al.}(2022)\citenamefont {Tusi}, \citenamefont {Franchi}, \citenamefont {Livi}, \citenamefont {Baumann}, \citenamefont {Benedicto~Orenes}, \citenamefont {Del~Re}, \citenamefont {Barfknecht}, \citenamefont {Zhou}, \citenamefont {Inguscio}, \citenamefont {Cappellini}, \citenamefont {Capone}, \citenamefont {Catani},\ and\ \citenamefont {Fallani}}]{Tusi2022}%
  \BibitemOpen
  \bibfield  {author} {\bibinfo {author} {\bibfnamefont {D.}~\bibnamefont {Tusi}}, \bibinfo {author} {\bibfnamefont {L.}~\bibnamefont {Franchi}}, \bibinfo {author} {\bibfnamefont {L.~F.}\ \bibnamefont {Livi}}, \bibinfo {author} {\bibfnamefont {K.}~\bibnamefont {Baumann}}, \bibinfo {author} {\bibfnamefont {D.}~\bibnamefont {Benedicto~Orenes}}, \bibinfo {author} {\bibfnamefont {L.}~\bibnamefont {Del~Re}}, \bibinfo {author} {\bibfnamefont {R.~E.}\ \bibnamefont {Barfknecht}}, \bibinfo {author} {\bibfnamefont {T.-W.}\ \bibnamefont {Zhou}}, \bibinfo {author} {\bibfnamefont {M.}~\bibnamefont {Inguscio}}, \bibinfo {author} {\bibfnamefont {G.}~\bibnamefont {Cappellini}}, \bibinfo {author} {\bibfnamefont {M.}~\bibnamefont {Capone}}, \bibinfo {author} {\bibfnamefont {J.}~\bibnamefont {Catani}},\ and\ \bibinfo {author} {\bibfnamefont {L.}~\bibnamefont {Fallani}},\ }\href {https://doi.org/10.1038/s41567-022-01726-5} {\bibfield  {journal} {\bibinfo  {journal} {Nat. Phys.}\ }\textbf {\bibinfo {volume} {18}},\ \bibinfo
  {pages} {1201} (\bibinfo {year} {2022})}\BibitemShut {NoStop}%
\bibitem [{\citenamefont {Capone}\ \emph {et~al.}(2002{\natexlab{a}})\citenamefont {Capone}, \citenamefont {Fabrizio}, \citenamefont {Castellani},\ and\ \citenamefont {Tosatti}}]{Capone_Science}%
  \BibitemOpen
  \bibfield  {author} {\bibinfo {author} {\bibfnamefont {M.}~\bibnamefont {Capone}}, \bibinfo {author} {\bibfnamefont {M.}~\bibnamefont {Fabrizio}}, \bibinfo {author} {\bibfnamefont {C.}~\bibnamefont {Castellani}},\ and\ \bibinfo {author} {\bibfnamefont {E.}~\bibnamefont {Tosatti}},\ }\href@noop {} {\bibfield  {journal} {\bibinfo  {journal} {Science}\ }\textbf {\bibinfo {volume} {296}},\ \bibinfo {pages} {2364} (\bibinfo {year} {2002}{\natexlab{a}})}\BibitemShut {NoStop}%
\bibitem [{\citenamefont {Capone}\ \emph {et~al.}(2010)\citenamefont {Capone}, \citenamefont {Castellani},\ and\ \citenamefont {Grilli}}]{Capone2010}%
  \BibitemOpen
  \bibfield  {author} {\bibinfo {author} {\bibfnamefont {M.}~\bibnamefont {Capone}}, \bibinfo {author} {\bibfnamefont {C.}~\bibnamefont {Castellani}},\ and\ \bibinfo {author} {\bibfnamefont {M.}~\bibnamefont {Grilli}},\ }\href {https://doi.org/https://doi.org/10.1155/2010/920860} {\bibfield  {journal} {\bibinfo  {journal} {Advances in Condensed Matter Physics}\ }\textbf {\bibinfo {volume} {2010}},\ \bibinfo {pages} {920860} (\bibinfo {year} {2010})}\BibitemShut {NoStop}%
\bibitem [{\citenamefont {Capone}\ \emph {et~al.}(2009)\citenamefont {Capone}, \citenamefont {Fabrizio}, \citenamefont {Castellani},\ and\ \citenamefont {Tosatti}}]{Capone2009}%
  \BibitemOpen
  \bibfield  {author} {\bibinfo {author} {\bibfnamefont {M.}~\bibnamefont {Capone}}, \bibinfo {author} {\bibfnamefont {M.}~\bibnamefont {Fabrizio}}, \bibinfo {author} {\bibfnamefont {C.}~\bibnamefont {Castellani}},\ and\ \bibinfo {author} {\bibfnamefont {E.}~\bibnamefont {Tosatti}},\ }\href {https://doi.org/10.1103/RevModPhys.81.943} {\bibfield  {journal} {\bibinfo  {journal} {Rev. Mod. Phys.}\ }\textbf {\bibinfo {volume} {81}},\ \bibinfo {pages} {943} (\bibinfo {year} {2009})}\BibitemShut {NoStop}%
\bibitem [{\citenamefont {Koga}\ and\ \citenamefont {Werner}(2015)}]{Koga2015}%
  \BibitemOpen
  \bibfield  {author} {\bibinfo {author} {\bibfnamefont {A.}~\bibnamefont {Koga}}\ and\ \bibinfo {author} {\bibfnamefont {P.}~\bibnamefont {Werner}},\ }\href {https://doi.org/10.1103/PhysRevB.91.085108} {\bibfield  {journal} {\bibinfo  {journal} {Phys. Rev. B}\ }\textbf {\bibinfo {volume} {91}},\ \bibinfo {pages} {085108} (\bibinfo {year} {2015})}\BibitemShut {NoStop}%
\bibitem [{\citenamefont {Zujev}\ \emph {et~al.}(2014)\citenamefont {Zujev}, \citenamefont {Scalettar}, \citenamefont {Batrouni},\ and\ \citenamefont {Sengupta}}]{Zujev_2014}%
  \BibitemOpen
  \bibfield  {author} {\bibinfo {author} {\bibfnamefont {A.}~\bibnamefont {Zujev}}, \bibinfo {author} {\bibfnamefont {R.~T.}\ \bibnamefont {Scalettar}}, \bibinfo {author} {\bibfnamefont {G.~G.}\ \bibnamefont {Batrouni}},\ and\ \bibinfo {author} {\bibfnamefont {P.}~\bibnamefont {Sengupta}},\ }\href {https://doi.org/10.1088/1367-2630/16/1/013004} {\bibfield  {journal} {\bibinfo  {journal} {New Journal of Physics}\ }\textbf {\bibinfo {volume} {16}},\ \bibinfo {pages} {013004} (\bibinfo {year} {2014})}\BibitemShut {NoStop}%
\bibitem [{\citenamefont {Koga}\ and\ \citenamefont {Werner}(2011)}]{Koga_2011}%
  \BibitemOpen
  \bibfield  {author} {\bibinfo {author} {\bibfnamefont {A.}~\bibnamefont {Koga}}\ and\ \bibinfo {author} {\bibfnamefont {P.}~\bibnamefont {Werner}},\ }\href {https://doi.org/10.1088/1742-6596/302/1/012040} {\bibfield  {journal} {\bibinfo  {journal} {Journal of Physics: Conference Series}\ }\textbf {\bibinfo {volume} {302}},\ \bibinfo {pages} {012040} (\bibinfo {year} {2011})}\BibitemShut {NoStop}%
\bibitem [{\citenamefont {Kamihara}\ \emph {et~al.}(2008)\citenamefont {Kamihara}, \citenamefont {Watanabe}, \citenamefont {Hirano},\ and\ \citenamefont {Hosono}}]{IBS2008}%
  \BibitemOpen
  \bibfield  {author} {\bibinfo {author} {\bibfnamefont {Y.}~\bibnamefont {Kamihara}}, \bibinfo {author} {\bibfnamefont {T.}~\bibnamefont {Watanabe}}, \bibinfo {author} {\bibfnamefont {M.}~\bibnamefont {Hirano}},\ and\ \bibinfo {author} {\bibfnamefont {H.}~\bibnamefont {Hosono}},\ }\href {https://doi.org/10.1021/ja800073m} {\bibfield  {journal} {\bibinfo  {journal} {Journal of the American Chemical Society}\ }\textbf {\bibinfo {volume} {130}},\ \bibinfo {pages} {3296} (\bibinfo {year} {2008})}\BibitemShut {NoStop}%
\bibitem [{\citenamefont {Fernandes}\ \emph {et~al.}(2022)\citenamefont {Fernandes}, \citenamefont {Coldea}, \citenamefont {Ding}, \citenamefont {Fisher}, \citenamefont {Hirschfeld},\ and\ \citenamefont {Kotliar}}]{Fernandes2022}%
  \BibitemOpen
  \bibfield  {author} {\bibinfo {author} {\bibfnamefont {R.~M.}\ \bibnamefont {Fernandes}}, \bibinfo {author} {\bibfnamefont {A.~I.}\ \bibnamefont {Coldea}}, \bibinfo {author} {\bibfnamefont {H.}~\bibnamefont {Ding}}, \bibinfo {author} {\bibfnamefont {I.~R.}\ \bibnamefont {Fisher}}, \bibinfo {author} {\bibfnamefont {P.~J.}\ \bibnamefont {Hirschfeld}},\ and\ \bibinfo {author} {\bibfnamefont {G.}~\bibnamefont {Kotliar}},\ }\href {https://doi.org/10.1038/s41586-021-04073-2} {\bibfield  {journal} {\bibinfo  {journal} {Nature}\ }\textbf {\bibinfo {volume} {601}},\ \bibinfo {pages} {35} (\bibinfo {year} {2022})}\BibitemShut {NoStop}%
\bibitem [{\citenamefont {Si}\ and\ \citenamefont {Hussey}(2023)}]{Si2023}%
  \BibitemOpen
  \bibfield  {author} {\bibinfo {author} {\bibfnamefont {Q.}~\bibnamefont {Si}}\ and\ \bibinfo {author} {\bibfnamefont {N.~E.}\ \bibnamefont {Hussey}},\ }\href {https://doi.org/10.1063/PT.3.5235} {\bibfield  {journal} {\bibinfo  {journal} {Physics Today}\ }\textbf {\bibinfo {volume} {76}},\ \bibinfo {pages} {34} (\bibinfo {year} {2023})},\ \Eprint {https://arxiv.org/abs/https://pubs.aip.org/physicstoday/article-pdf/76/5/34/20086106/34\_1\_pt.3.5235.pdf} {https://pubs.aip.org/physicstoday/article-pdf/76/5/34/20086106/34\_1\_pt.3.5235.pdf} \BibitemShut {NoStop}%
\bibitem [{\citenamefont {Haule}\ and\ \citenamefont {Kotliar}(2009)}]{Haule_Kotliar}%
  \BibitemOpen
  \bibfield  {author} {\bibinfo {author} {\bibfnamefont {K.}~\bibnamefont {Haule}}\ and\ \bibinfo {author} {\bibfnamefont {G.}~\bibnamefont {Kotliar}},\ }\href {https://doi.org/10.1088/1367-2630/11/2/025021} {\bibfield  {journal} {\bibinfo  {journal} {New Journal of Physics}\ }\textbf {\bibinfo {volume} {11}},\ \bibinfo {pages} {025021} (\bibinfo {year} {2009})}\BibitemShut {NoStop}%
\bibitem [{\citenamefont {de' Medici}(2011)}]{deMedici2011}%
  \BibitemOpen
  \bibfield  {author} {\bibinfo {author} {\bibfnamefont {L.}~\bibnamefont {de' Medici}},\ }\href {https://doi.org/10.1103/PhysRevB.83.205112} {\bibfield  {journal} {\bibinfo  {journal} {Phys. Rev. B}\ }\textbf {\bibinfo {volume} {83}},\ \bibinfo {pages} {205112} (\bibinfo {year} {2011})}\BibitemShut {NoStop}%
\bibitem [{\citenamefont {Georges}\ and\ \citenamefont {Kotliar}(2024)}]{Georges2024}%
  \BibitemOpen
  \bibfield  {author} {\bibinfo {author} {\bibfnamefont {A.}~\bibnamefont {Georges}}\ and\ \bibinfo {author} {\bibfnamefont {G.}~\bibnamefont {Kotliar}},\ }\href {https://doi.org/10.1063/pt.wqrz.qpjx} {\bibfield  {journal} {\bibinfo  {journal} {Physics Today}\ }\textbf {\bibinfo {volume} {77}},\ \bibinfo {pages} {46} (\bibinfo {year} {2024})},\ \Eprint {https://arxiv.org/abs/https://pubs.aip.org/physicstoday/article-pdf/77/4/46/20086423/46\_1\_pt.wqrz.qpjx.pdf} {https://pubs.aip.org/physicstoday/article-pdf/77/4/46/20086423/46\_1\_pt.wqrz.qpjx.pdf} \BibitemShut {NoStop}%
\bibitem [{\citenamefont {Yu}\ \emph {et~al.}(2021)\citenamefont {Yu}, \citenamefont {Hu}, \citenamefont {Nica}, \citenamefont {Zhu},\ and\ \citenamefont {Si}}]{Yu2021}%
  \BibitemOpen
  \bibfield  {author} {\bibinfo {author} {\bibfnamefont {R.}~\bibnamefont {Yu}}, \bibinfo {author} {\bibfnamefont {H.}~\bibnamefont {Hu}}, \bibinfo {author} {\bibfnamefont {E.~M.}\ \bibnamefont {Nica}}, \bibinfo {author} {\bibfnamefont {J.-X.}\ \bibnamefont {Zhu}},\ and\ \bibinfo {author} {\bibfnamefont {Q.}~\bibnamefont {Si}},\ }\bibfield  {journal} {\bibinfo  {journal} {Frontiers in Physics}\ }\textbf {\bibinfo {volume} {Volume 9 - 2021}},\ \href {https://doi.org/10.3389/fphy.2021.578347} {10.3389/fphy.2021.578347} (\bibinfo {year} {2021})\BibitemShut {NoStop}%
\bibitem [{\citenamefont {de' Medici}\ \emph {et~al.}(2014)\citenamefont {de' Medici}, \citenamefont {Giovannetti},\ and\ \citenamefont {Capone}}]{De_Medici_Selective_Mott}%
  \BibitemOpen
  \bibfield  {author} {\bibinfo {author} {\bibfnamefont {L.}~\bibnamefont {de' Medici}}, \bibinfo {author} {\bibfnamefont {G.}~\bibnamefont {Giovannetti}},\ and\ \bibinfo {author} {\bibfnamefont {M.}~\bibnamefont {Capone}},\ }\href {https://doi.org/10.1103/PhysRevLett.112.177001} {\bibfield  {journal} {\bibinfo  {journal} {Phys. Rev. Lett.}\ }\textbf {\bibinfo {volume} {112}},\ \bibinfo {pages} {177001} (\bibinfo {year} {2014})}\BibitemShut {NoStop}%
\bibitem [{\citenamefont {Nourafkan}\ \emph {et~al.}(2016)\citenamefont {Nourafkan}, \citenamefont {Kotliar},\ and\ \citenamefont {Tremblay}}]{Nourafkan2016}%
  \BibitemOpen
  \bibfield  {author} {\bibinfo {author} {\bibfnamefont {R.}~\bibnamefont {Nourafkan}}, \bibinfo {author} {\bibfnamefont {G.}~\bibnamefont {Kotliar}},\ and\ \bibinfo {author} {\bibfnamefont {A.-M.~S.}\ \bibnamefont {Tremblay}},\ }\href {https://doi.org/10.1103/PhysRevLett.117.137001} {\bibfield  {journal} {\bibinfo  {journal} {Phys. Rev. Lett.}\ }\textbf {\bibinfo {volume} {117}},\ \bibinfo {pages} {137001} (\bibinfo {year} {2016})}\BibitemShut {NoStop}%
\bibitem [{\citenamefont {Ido}\ \emph {et~al.}(2023)\citenamefont {Ido}, \citenamefont {Motoyama}, \citenamefont {Yoshimi},\ and\ \citenamefont {Misawa}}]{Ido2023}%
  \BibitemOpen
  \bibfield  {author} {\bibinfo {author} {\bibfnamefont {K.}~\bibnamefont {Ido}}, \bibinfo {author} {\bibfnamefont {Y.}~\bibnamefont {Motoyama}}, \bibinfo {author} {\bibfnamefont {K.}~\bibnamefont {Yoshimi}},\ and\ \bibinfo {author} {\bibfnamefont {T.}~\bibnamefont {Misawa}},\ }\href {https://doi.org/10.7566/JPSJ.92.064702} {\bibfield  {journal} {\bibinfo  {journal} {Journal of the Physical Society of Japan}\ }\textbf {\bibinfo {volume} {92}},\ \bibinfo {pages} {064702} (\bibinfo {year} {2023})},\ \Eprint {https://arxiv.org/abs/https://doi.org/10.7566/JPSJ.92.064702} {https://doi.org/10.7566/JPSJ.92.064702} \BibitemShut {NoStop}%
\bibitem [{\citenamefont {Fanfarillo}\ \emph {et~al.}(2020)\citenamefont {Fanfarillo}, \citenamefont {Valli},\ and\ \citenamefont {Capone}}]{Fanfarillo2020}%
  \BibitemOpen
  \bibfield  {author} {\bibinfo {author} {\bibfnamefont {L.}~\bibnamefont {Fanfarillo}}, \bibinfo {author} {\bibfnamefont {A.}~\bibnamefont {Valli}},\ and\ \bibinfo {author} {\bibfnamefont {M.}~\bibnamefont {Capone}},\ }\href {https://doi.org/10.1103/PhysRevLett.125.177001} {\bibfield  {journal} {\bibinfo  {journal} {Phys. Rev. Lett.}\ }\textbf {\bibinfo {volume} {125}},\ \bibinfo {pages} {177001} (\bibinfo {year} {2020})}\BibitemShut {NoStop}%
\bibitem [{\citenamefont {Marino}\ \emph {et~al.}(2024)\citenamefont {Marino}, \citenamefont {Scazzola}, \citenamefont {Becca}, \citenamefont {Capone},\ and\ \citenamefont {L.F.}}]{Vito}%
  \BibitemOpen
  \bibfield  {author} {\bibinfo {author} {\bibfnamefont {V.}~\bibnamefont {Marino}}, \bibinfo {author} {\bibfnamefont {A.}~\bibnamefont {Scazzola}}, \bibinfo {author} {\bibfnamefont {F.}~\bibnamefont {Becca}}, \bibinfo {author} {\bibfnamefont {M.}~\bibnamefont {Capone}},\ and\ \bibinfo {author} {\bibfnamefont {T.}~\bibnamefont {L.F.}},\ }\href@noop {} {\bibinfo {title} {Intertwined superconductivity and orbital selectivity in a three-orbital hubbard model for the iron pnictides}} (\bibinfo {year} {2024}),\ \Eprint {https://arxiv.org/abs/arXiv:2406.13634} {arXiv:2406.13634} \BibitemShut {NoStop}%
\bibitem [{\citenamefont {Boeri}\ \emph {et~al.}(2008)\citenamefont {Boeri}, \citenamefont {Dolgov},\ and\ \citenamefont {Golubov}}]{Lilia}%
  \BibitemOpen
  \bibfield  {author} {\bibinfo {author} {\bibfnamefont {L.}~\bibnamefont {Boeri}}, \bibinfo {author} {\bibfnamefont {O.~V.}\ \bibnamefont {Dolgov}},\ and\ \bibinfo {author} {\bibfnamefont {A.~A.}\ \bibnamefont {Golubov}},\ }\href {https://doi.org/10.1103/PhysRevLett.101.026403} {\bibfield  {journal} {\bibinfo  {journal} {Phys. Rev. Lett.}\ }\textbf {\bibinfo {volume} {101}},\ \bibinfo {pages} {026403} (\bibinfo {year} {2008})}\BibitemShut {NoStop}%
\bibitem [{\citenamefont {Zhang}\ \emph {et~al.}(2017)\citenamefont {Zhang}, \citenamefont {Liu}, \citenamefont {Chen}, \citenamefont {Xie}, \citenamefont {He}, \citenamefont {Tang}, \citenamefont {He}, \citenamefont {Li}, \citenamefont {Jia}, \citenamefont {Rebec}, \citenamefont {Ma}, \citenamefont {Yan}, \citenamefont {Hashimoto}, \citenamefont {Lu}, \citenamefont {Mo}, \citenamefont {Hikita}, \citenamefont {Moore}, \citenamefont {Hwang}, \citenamefont {Lee},\ and\ \citenamefont {Shen}}]{Zhang_2017}%
  \BibitemOpen
  \bibfield  {author} {\bibinfo {author} {\bibfnamefont {C.}~\bibnamefont {Zhang}}, \bibinfo {author} {\bibfnamefont {Z.}~\bibnamefont {Liu}}, \bibinfo {author} {\bibfnamefont {Z.}~\bibnamefont {Chen}}, \bibinfo {author} {\bibfnamefont {Y.}~\bibnamefont {Xie}}, \bibinfo {author} {\bibfnamefont {R.}~\bibnamefont {He}}, \bibinfo {author} {\bibfnamefont {S.}~\bibnamefont {Tang}}, \bibinfo {author} {\bibfnamefont {J.}~\bibnamefont {He}}, \bibinfo {author} {\bibfnamefont {W.}~\bibnamefont {Li}}, \bibinfo {author} {\bibfnamefont {T.}~\bibnamefont {Jia}}, \bibinfo {author} {\bibfnamefont {S.~N.}\ \bibnamefont {Rebec}}, \bibinfo {author} {\bibfnamefont {E.~Y.}\ \bibnamefont {Ma}}, \bibinfo {author} {\bibfnamefont {H.}~\bibnamefont {Yan}}, \bibinfo {author} {\bibfnamefont {M.}~\bibnamefont {Hashimoto}}, \bibinfo {author} {\bibfnamefont {D.}~\bibnamefont {Lu}}, \bibinfo {author} {\bibfnamefont {S.-K.}\ \bibnamefont {Mo}}, \bibinfo {author} {\bibfnamefont {Y.}~\bibnamefont {Hikita}}, \bibinfo {author} {\bibfnamefont
  {R.~G.}\ \bibnamefont {Moore}}, \bibinfo {author} {\bibfnamefont {H.~Y.}\ \bibnamefont {Hwang}}, \bibinfo {author} {\bibfnamefont {D.}~\bibnamefont {Lee}},\ and\ \bibinfo {author} {\bibfnamefont {Z.}~\bibnamefont {Shen}},\ }\bibfield  {journal} {\bibinfo  {journal} {Nature Communications}\ }\textbf {\bibinfo {volume} {8}},\ \href {https://doi.org/10.1038/ncomms14468} {10.1038/ncomms14468} (\bibinfo {year} {2017})\BibitemShut {NoStop}%
\bibitem [{\citenamefont {Zhang}\ \emph {et~al.}(2015)\citenamefont {Zhang}, \citenamefont {Wang}, \citenamefont {Song}, \citenamefont {Liu}, \citenamefont {Peng}, \citenamefont {Moler}, \citenamefont {Feng},\ and\ \citenamefont {Wang}}]{Zhang_2015}%
  \BibitemOpen
  \bibfield  {author} {\bibinfo {author} {\bibfnamefont {Z.}~\bibnamefont {Zhang}}, \bibinfo {author} {\bibfnamefont {Y.-H.}\ \bibnamefont {Wang}}, \bibinfo {author} {\bibfnamefont {Q.}~\bibnamefont {Song}}, \bibinfo {author} {\bibfnamefont {C.}~\bibnamefont {Liu}}, \bibinfo {author} {\bibfnamefont {R.}~\bibnamefont {Peng}}, \bibinfo {author} {\bibfnamefont {K.}~\bibnamefont {Moler}}, \bibinfo {author} {\bibfnamefont {D.}~\bibnamefont {Feng}},\ and\ \bibinfo {author} {\bibfnamefont {Y.}~\bibnamefont {Wang}},\ }\href {https://doi.org/10.1007/s11434-015-0842-8} {\bibfield  {journal} {\bibinfo  {journal} {Science Bulletin}\ }\textbf {\bibinfo {volume} {60}},\ \bibinfo {pages} {1301–1304} (\bibinfo {year} {2015})}\BibitemShut {NoStop}%
\bibitem [{\citenamefont {Mandal}\ \emph {et~al.}(2014)\citenamefont {Mandal}, \citenamefont {Cohen},\ and\ \citenamefont {Haule}}]{Mandal2014}%
  \BibitemOpen
  \bibfield  {author} {\bibinfo {author} {\bibfnamefont {S.}~\bibnamefont {Mandal}}, \bibinfo {author} {\bibfnamefont {R.~E.}\ \bibnamefont {Cohen}},\ and\ \bibinfo {author} {\bibfnamefont {K.}~\bibnamefont {Haule}},\ }\href {https://doi.org/10.1103/PhysRevB.89.220502} {\bibfield  {journal} {\bibinfo  {journal} {Phys. Rev. B}\ }\textbf {\bibinfo {volume} {89}},\ \bibinfo {pages} {220502} (\bibinfo {year} {2014})}\BibitemShut {NoStop}%
\bibitem [{\citenamefont {Choi}\ \emph {et~al.}(2017)\citenamefont {Choi}, \citenamefont {Johnston}, \citenamefont {Jang}, \citenamefont {Koepernik}, \citenamefont {Nakatsukasa}, \citenamefont {Ok}, \citenamefont {Lee}, \citenamefont {Choi}, \citenamefont {Lee}, \citenamefont {Akbari}, \citenamefont {Semertzidis}, \citenamefont {Bang}, \citenamefont {Kim},\ and\ \citenamefont {Lee}}]{PhysRevLett.119.107003}%
  \BibitemOpen
  \bibfield  {author} {\bibinfo {author} {\bibfnamefont {S.}~\bibnamefont {Choi}}, \bibinfo {author} {\bibfnamefont {S.}~\bibnamefont {Johnston}}, \bibinfo {author} {\bibfnamefont {W.-J.}\ \bibnamefont {Jang}}, \bibinfo {author} {\bibfnamefont {K.}~\bibnamefont {Koepernik}}, \bibinfo {author} {\bibfnamefont {K.}~\bibnamefont {Nakatsukasa}}, \bibinfo {author} {\bibfnamefont {J.~M.}\ \bibnamefont {Ok}}, \bibinfo {author} {\bibfnamefont {H.-J.}\ \bibnamefont {Lee}}, \bibinfo {author} {\bibfnamefont {H.~W.}\ \bibnamefont {Choi}}, \bibinfo {author} {\bibfnamefont {A.~T.}\ \bibnamefont {Lee}}, \bibinfo {author} {\bibfnamefont {A.}~\bibnamefont {Akbari}}, \bibinfo {author} {\bibfnamefont {Y.~K.}\ \bibnamefont {Semertzidis}}, \bibinfo {author} {\bibfnamefont {Y.}~\bibnamefont {Bang}}, \bibinfo {author} {\bibfnamefont {J.~S.}\ \bibnamefont {Kim}},\ and\ \bibinfo {author} {\bibfnamefont {J.}~\bibnamefont {Lee}},\ }\href {https://doi.org/10.1103/PhysRevLett.119.107003} {\bibfield  {journal} {\bibinfo  {journal} {Phys.
  Rev. Lett.}\ }\textbf {\bibinfo {volume} {119}},\ \bibinfo {pages} {107003} (\bibinfo {year} {2017})}\BibitemShut {NoStop}%
\bibitem [{\citenamefont {Capone}\ \emph {et~al.}(2001)\citenamefont {Capone}, \citenamefont {Fabrizio},\ and\ \citenamefont {Tosatti}}]{Capone2001}%
  \BibitemOpen
  \bibfield  {author} {\bibinfo {author} {\bibfnamefont {M.}~\bibnamefont {Capone}}, \bibinfo {author} {\bibfnamefont {M.}~\bibnamefont {Fabrizio}},\ and\ \bibinfo {author} {\bibfnamefont {E.}~\bibnamefont {Tosatti}},\ }\href {https://doi.org/10.1103/PhysRevLett.86.5361} {\bibfield  {journal} {\bibinfo  {journal} {Phys. Rev. Lett.}\ }\textbf {\bibinfo {volume} {86}},\ \bibinfo {pages} {5361} (\bibinfo {year} {2001})}\BibitemShut {NoStop}%
\bibitem [{\citenamefont {Scazzola}\ \emph {et~al.}(2023)\citenamefont {Scazzola}, \citenamefont {Amaricci},\ and\ \citenamefont {Capone}}]{Scazzola2023}%
  \BibitemOpen
  \bibfield  {author} {\bibinfo {author} {\bibfnamefont {A.}~\bibnamefont {Scazzola}}, \bibinfo {author} {\bibfnamefont {A.}~\bibnamefont {Amaricci}},\ and\ \bibinfo {author} {\bibfnamefont {M.}~\bibnamefont {Capone}},\ }\href {https://doi.org/10.1103/PhysRevB.107.085131} {\bibfield  {journal} {\bibinfo  {journal} {Phys. Rev. B}\ }\textbf {\bibinfo {volume} {107}},\ \bibinfo {pages} {085131} (\bibinfo {year} {2023})}\BibitemShut {NoStop}%
\bibitem [{\citenamefont {Nomura}\ \emph {et~al.}(01x5)\citenamefont {Nomura}, \citenamefont {Sakai}, \citenamefont {Capone},\ and\ \citenamefont {Arita}}]{Nomura_2015}%
  \BibitemOpen
  \bibfield  {author} {\bibinfo {author} {\bibfnamefont {Y.}~\bibnamefont {Nomura}}, \bibinfo {author} {\bibfnamefont {S.}~\bibnamefont {Sakai}}, \bibinfo {author} {\bibfnamefont {M.}~\bibnamefont {Capone}},\ and\ \bibinfo {author} {\bibfnamefont {R.}~\bibnamefont {Arita}},\ }\bibfield  {journal} {\bibinfo  {journal} {Science Advances}\ }\textbf {\bibinfo {volume} {1}},\ \href {https://doi.org/10.1126/sciadv.1500568} {10.1126/sciadv.1500568} (\bibinfo {year} {201x5})\BibitemShut {NoStop}%
\bibitem [{\citenamefont {Sangiovanni}\ \emph {et~al.}(2006)\citenamefont {Sangiovanni}, \citenamefont {Capone},\ and\ \citenamefont {Castellani}}]{Sangiovanni2006}%
  \BibitemOpen
  \bibfield  {author} {\bibinfo {author} {\bibfnamefont {G.}~\bibnamefont {Sangiovanni}}, \bibinfo {author} {\bibfnamefont {M.}~\bibnamefont {Capone}},\ and\ \bibinfo {author} {\bibfnamefont {C.}~\bibnamefont {Castellani}},\ }\href {https://doi.org/10.1103/PhysRevB.73.165123} {\bibfield  {journal} {\bibinfo  {journal} {Phys. Rev. B}\ }\textbf {\bibinfo {volume} {73}},\ \bibinfo {pages} {165123} (\bibinfo {year} {2006})}\BibitemShut {NoStop}%
\bibitem [{\citenamefont {Georges}\ \emph {et~al.}(1996)\citenamefont {Georges}, \citenamefont {Kotliar}, \citenamefont {Krauth},\ and\ \citenamefont {Rozenberg}}]{Georges}%
  \BibitemOpen
  \bibfield  {author} {\bibinfo {author} {\bibfnamefont {A.}~\bibnamefont {Georges}}, \bibinfo {author} {\bibfnamefont {G.}~\bibnamefont {Kotliar}}, \bibinfo {author} {\bibfnamefont {W.}~\bibnamefont {Krauth}},\ and\ \bibinfo {author} {\bibfnamefont {M.~J.}\ \bibnamefont {Rozenberg}},\ }\href {https://doi.org/10.1103/RevModPhys.68.13} {\bibfield  {journal} {\bibinfo  {journal} {Rev. Mod. Phys.}\ }\textbf {\bibinfo {volume} {68}},\ \bibinfo {pages} {13} (\bibinfo {year} {1996})}\BibitemShut {NoStop}%
\bibitem [{\citenamefont {Caffarel}\ and\ \citenamefont {Krauth}(1994)}]{Caffarel1994}%
  \BibitemOpen
  \bibfield  {author} {\bibinfo {author} {\bibfnamefont {M.}~\bibnamefont {Caffarel}}\ and\ \bibinfo {author} {\bibfnamefont {W.}~\bibnamefont {Krauth}},\ }\href {https://doi.org/10.1103/PhysRevLett.72.1545} {\bibfield  {journal} {\bibinfo  {journal} {Phys. Rev. Lett.}\ }\textbf {\bibinfo {volume} {72}},\ \bibinfo {pages} {1545} (\bibinfo {year} {1994})}\BibitemShut {NoStop}%
\bibitem [{\citenamefont {Capone}\ \emph {et~al.}(2007)\citenamefont {Capone}, \citenamefont {de' Medici},\ and\ \citenamefont {Georges}}]{Capone2007}%
  \BibitemOpen
  \bibfield  {author} {\bibinfo {author} {\bibfnamefont {M.}~\bibnamefont {Capone}}, \bibinfo {author} {\bibfnamefont {L.}~\bibnamefont {de' Medici}},\ and\ \bibinfo {author} {\bibfnamefont {A.}~\bibnamefont {Georges}},\ }\href {https://doi.org/10.1103/PhysRevB.76.245116} {\bibfield  {journal} {\bibinfo  {journal} {Phys. Rev. B}\ }\textbf {\bibinfo {volume} {76}},\ \bibinfo {pages} {245116} (\bibinfo {year} {2007})}\BibitemShut {NoStop}%
\bibitem [{\citenamefont {Amaricci}\ \emph {et~al.}(2022)\citenamefont {Amaricci}, \citenamefont {Crippa}, \citenamefont {Scazzola}, \citenamefont {Petocchi}, \citenamefont {Mazza}, \citenamefont {{de Medici}},\ and\ \citenamefont {Capone}}]{AMARICCI2022108261}%
  \BibitemOpen
  \bibfield  {author} {\bibinfo {author} {\bibfnamefont {A.}~\bibnamefont {Amaricci}}, \bibinfo {author} {\bibfnamefont {L.}~\bibnamefont {Crippa}}, \bibinfo {author} {\bibfnamefont {A.}~\bibnamefont {Scazzola}}, \bibinfo {author} {\bibfnamefont {F.}~\bibnamefont {Petocchi}}, \bibinfo {author} {\bibfnamefont {G.}~\bibnamefont {Mazza}}, \bibinfo {author} {\bibfnamefont {L.}~\bibnamefont {{de Medici}}},\ and\ \bibinfo {author} {\bibfnamefont {M.}~\bibnamefont {Capone}},\ }\href {https://doi.org/https://doi.org/10.1016/j.cpc.2021.108261} {\bibfield  {journal} {\bibinfo  {journal} {Computer Physics Communications}\ }\textbf {\bibinfo {volume} {273}},\ \bibinfo {pages} {108261} (\bibinfo {year} {2022})}\BibitemShut {NoStop}%
\bibitem [{\citenamefont {Toschi}\ \emph {et~al.}(2005{\natexlab{a}})\citenamefont {Toschi}, \citenamefont {Capone},\ and\ \citenamefont {Castellani}}]{Toschi_BCSBEC}%
  \BibitemOpen
  \bibfield  {author} {\bibinfo {author} {\bibfnamefont {A.}~\bibnamefont {Toschi}}, \bibinfo {author} {\bibfnamefont {M.}~\bibnamefont {Capone}},\ and\ \bibinfo {author} {\bibfnamefont {C.}~\bibnamefont {Castellani}},\ }\href {https://doi.org/10.1103/PhysRevB.72.235118} {\bibfield  {journal} {\bibinfo  {journal} {Phys. Rev. B}\ }\textbf {\bibinfo {volume} {72}},\ \bibinfo {pages} {235118} (\bibinfo {year} {2005}{\natexlab{a}})}\BibitemShut {NoStop}%
\bibitem [{\citenamefont {Kyung}\ \emph {et~al.}(2006)\citenamefont {Kyung}, \citenamefont {Georges},\ and\ \citenamefont {Tremblay}}]{Kyung2006}%
  \BibitemOpen
  \bibfield  {author} {\bibinfo {author} {\bibfnamefont {B.}~\bibnamefont {Kyung}}, \bibinfo {author} {\bibfnamefont {A.}~\bibnamefont {Georges}},\ and\ \bibinfo {author} {\bibfnamefont {A.-M.~S.}\ \bibnamefont {Tremblay}},\ }\href {https://doi.org/10.1103/PhysRevB.74.024501} {\bibfield  {journal} {\bibinfo  {journal} {Phys. Rev. B}\ }\textbf {\bibinfo {volume} {74}},\ \bibinfo {pages} {024501} (\bibinfo {year} {2006})}\BibitemShut {NoStop}%
\bibitem [{\citenamefont {Toschi}\ \emph {et~al.}(2005{\natexlab{b}})\citenamefont {Toschi}, \citenamefont {Barone}, \citenamefont {Capone},\ and\ \citenamefont {Castellani}}]{Toschi_2005}%
  \BibitemOpen
  \bibfield  {author} {\bibinfo {author} {\bibfnamefont {A.}~\bibnamefont {Toschi}}, \bibinfo {author} {\bibfnamefont {P.}~\bibnamefont {Barone}}, \bibinfo {author} {\bibfnamefont {M.}~\bibnamefont {Capone}},\ and\ \bibinfo {author} {\bibfnamefont {C.}~\bibnamefont {Castellani}},\ }\href {https://doi.org/10.1088/1367-2630/7/1/007} {\bibfield  {journal} {\bibinfo  {journal} {New Journal of Physics}\ }\textbf {\bibinfo {volume} {7}},\ \bibinfo {pages} {7} (\bibinfo {year} {2005}{\natexlab{b}})}\BibitemShut {NoStop}%
\bibitem [{\citenamefont {Keller}\ \emph {et~al.}(2001)\citenamefont {Keller}, \citenamefont {Metzner},\ and\ \citenamefont {Schollw\"ock}}]{Keller2001}%
  \BibitemOpen
  \bibfield  {author} {\bibinfo {author} {\bibfnamefont {M.}~\bibnamefont {Keller}}, \bibinfo {author} {\bibfnamefont {W.}~\bibnamefont {Metzner}},\ and\ \bibinfo {author} {\bibfnamefont {U.}~\bibnamefont {Schollw\"ock}},\ }\href {https://doi.org/10.1103/PhysRevLett.86.4612} {\bibfield  {journal} {\bibinfo  {journal} {Phys. Rev. Lett.}\ }\textbf {\bibinfo {volume} {86}},\ \bibinfo {pages} {4612} (\bibinfo {year} {2001})}\BibitemShut {NoStop}%
\bibitem [{\citenamefont {Capone}\ \emph {et~al.}(2002{\natexlab{b}})\citenamefont {Capone}, \citenamefont {Castellani},\ and\ \citenamefont {Grilli}}]{Capone2002}%
  \BibitemOpen
  \bibfield  {author} {\bibinfo {author} {\bibfnamefont {M.}~\bibnamefont {Capone}}, \bibinfo {author} {\bibfnamefont {C.}~\bibnamefont {Castellani}},\ and\ \bibinfo {author} {\bibfnamefont {M.}~\bibnamefont {Grilli}},\ }\href {https://doi.org/10.1103/PhysRevLett.88.126403} {\bibfield  {journal} {\bibinfo  {journal} {Phys. Rev. Lett.}\ }\textbf {\bibinfo {volume} {88}},\ \bibinfo {pages} {126403} (\bibinfo {year} {2002}{\natexlab{b}})}\BibitemShut {NoStop}%
\bibitem [{\citenamefont {Witt}\ \emph {et~al.}(2024)\citenamefont {Witt}, \citenamefont {Nomura}, \citenamefont {Brener}, \citenamefont {Arita}, \citenamefont {Lichtenstein},\ and\ \citenamefont {Wehling}}]{Witt2024}%
  \BibitemOpen
  \bibfield  {author} {\bibinfo {author} {\bibfnamefont {N.}~\bibnamefont {Witt}}, \bibinfo {author} {\bibfnamefont {Y.}~\bibnamefont {Nomura}}, \bibinfo {author} {\bibfnamefont {S.}~\bibnamefont {Brener}}, \bibinfo {author} {\bibfnamefont {R.}~\bibnamefont {Arita}}, \bibinfo {author} {\bibfnamefont {A.~I.}\ \bibnamefont {Lichtenstein}},\ and\ \bibinfo {author} {\bibfnamefont {T.~O.}\ \bibnamefont {Wehling}},\ }\href {https://doi.org/10.1038/s41535-024-00706-7} {\bibfield  {journal} {\bibinfo  {journal} {npj Quantum Materials}\ }\textbf {\bibinfo {volume} {9}},\ \bibinfo {pages} {100} (\bibinfo {year} {2024})}\BibitemShut {NoStop}%
\end{thebibliography}%

\end{document}